\documentstyle[11pt]{article}
\parskip =\medskipamount
\def\be{\begin{equation}}
\def\ee{\end{equation}}
\def\disp{\displaystyle}

\newcounter{fig}

\def\Z{\rm {\sf Z\hspace{-3.15mm}Z\hspace{0.6mm}}}

\begin{document}

\begin{titlepage}

\begin{center}

{\LARGE \bf Localization in Simple Multiparticle \\ Catalytic Absorption
Model

}
\vspace{0.6in}

{\large Sergei Nechaev$^{\dag}$}
\bigskip

{\sl
Institut de Physique Nucl\'eaire, Division de Physique Th\'eorique$^{*}$,
\\ 91406 Orsay Cedex, France

and

L.D. Landau Institute for Theoretical Physics, \\ 117940, Moscow, Russia

}

\end{center}

\vspace{1in}

\begin{abstract}
We consider the phase transition in the system of $n$ simultaneously
developing random walks on the halfline $x\ge 0$. All walks are independent
on each others in all points except the origin $x=0$, where the point
well is located. The well depth depends on the number of particles (i.e. 
steps of different random walks) simultaneously staying at the point $x=0$. 
We consider the limit $n\gg 1$ and show that if the well depth grows faster 
than $\frac{3}{2}n\ln n$ with $n$, then all random walks become localized 
simultaneously at the origin. In the conclusion we discuss the connection 
of the above defined problem with the phase transition in the copolymer 
chain with quenched random sequence of monomers considered in the 
frameworks of replica approach. 
\end{abstract} 
\vspace{0.4in}

\vspace{0.3in}

\noindent {\sl Submitted to J. Phys.(A): Math. \& Gen.}
\vspace{0.3in}

\vspace{1.2in}

\hrule \footnotesize
\noindent $^{\dag}$ E-mail: {\it nechaev@landau.ac.ru} \\
$^{*}$ Unit\'e de Recherche des Universit\'es Paris XI et Paris VI
associ\'ee au C.N.R.S.

\end{titlepage}

{
\small
\tableofcontents
}

\section*{Introduction}

Problems dealing with the localization of random walks in potentials of
various geometries compel much attention of different groups of scientists:
both chemists and physicists. For the first group these problems are naturally
connected with creating a new materials possessing specific technological
properties (for instance, catalysts \cite{catal} or ion containing and 
surface active substances \cite{pav}). Meanwhile, for physicists the 
consideration of random walks localization remains since 70's the firing 
ground for elaborating the new methods of the investigation of a polymer 
absorption in different geometries, wetting phenomena and kinetics of 
chemical reactions.

The phenomenon of polymer chain absorption is rather well understood at
present both for homo-- \cite{bin}--\cite{bir3}, \cite{gin} and
heteropolymer \cite{gro2}--\cite{orl} cases. The simple diffusive approach
\cite{bir2,bir3,gin} provides complete understanding of an absorption of
homopolymer chains as well as of block--copolymers in complicated
geometries. These results can be beautifully explained also by
scaling estimates \cite{gen}. More advanced renormalization group methods
\cite{gro2} and power series analysis \cite{obu} applied to random chains
with a disordered sequence of links are also widely used and give an
exhaustive information about the thermodynamic properties of ideal polymers
near the point of phase transition from the delocalized (Gaussian) to the
localized (absorbed) fluctuation regimes. The similar mathematical 
formalism has been applied to a description of a phase transition in 
solid--on--solid (SOS) models with quenched impurities \cite{forg}. Many 
conclusions obtained by RG analysis in \cite{gro2} correlate with the 
results in \cite{forg}.

Problems dealing with 2D wetting subject to a layer of periodic potential
could be attributed to this theme as well. In \cite{zhang} using a
generalized transfer matrix method, we were able to find the exact solution
for the critical depinning phase transition.

Finally, it is worthwhile to mention that the model considered below
has some features of so-called "zero's range process" \cite{krug,evans}
emerged in connection with traffic models. Also, similar set of problems
appeared in course of investigation of the "autolocalization" phenomena
\cite{gred} and in the theory of multicomponent chemical reactions on traps
immersed in an immovable "matrix" \cite{osha,cardy}.

The present paper is organised as follows. In Section \ref{sec:1} we
formulate the simple model of "catalytic absorption" and put forward the
question of our main interest; Section \ref{sec:2} is devoted to the
solution of one-- and two--particle problems; the consideration of the 
phase transition in the $n$--particle model for $n>2$ is given in Section
\ref{sec:3}; while in Section \ref{sec:4} we present conjectures about the
connection of the multiparticle catalytic absorption with the random
copolymer absorption at the point well.

\section{Multiparticle model of catalytic absorption at the point well}
\label{sec:1}
\setcounter{equation}{0}

Consider the one--dimensional lattice ${\Z}^{+}$, i.e. the set of integer
points on the halfline $x\ge 0$, $\{x\}=\{0,\,1,\,2,\,3,\,\dots\}$. Take
$n$ independent $N$--step simplest random walks on ${\Z}^{+}$,
simultaneously starting at the point $x=0$. All these walks are free, i.e.
they do not interact neither with any external potential, nor with each
others in all points except the point $x=0$. At the point $x=0$ the
interaction of a given random walk with the well is described by the
potential $U(x_1,x_2,\ldots,x_n)$ (the index $i\in[1,n]$ lables the 
different random walks):
\be \label{eq:1}
U\left(x_1,x_2,\ldots,x_n\right)=\left\{
\begin{array}{ll}
0 & \mbox{if $x_i\neq 0$ for all $i\in [1,n]$} \\
U^{(1)} & \mbox{if $x_i=0;\; x_j\neq 0$ for all $\left\{j\neq i;\;
j\in [1,n] \right\}$} \\
U^{(2)} & \mbox{if $x_i=0,\; x_j=0; \; x_k\neq 0$ for all $\left\{k\neq
j\neq i;\; k\in [1,n] \right\}$} \\
\cdots & \\
U^{(n)} & \mbox{if $x_1=0,\; x_2=0,\; \ldots,\; x_n=0$}
\end{array} \right.
\ee
where
\be \label{eq:1.1}
0\le U^{(1)}\le U^{(2)}\le \cdots \le U^{(n)}
\ee
and
\be \label{eq:2}
U^{(i)}\equiv f(i)\,;\qquad i\in [1,n]
\ee

The equations (\ref{eq:1})--(\ref{eq:2}) reflect the fact that the depth
of the potential well depends on the number of particles (i.e. steps of 
different random walks) {\it simultaneously} located at the point 
$x=0$---see Fig.\ref{fig:1}. Let us stress that $U^{(i)}$ is the 
"individual" potential, i.e. the potential per one particle in a cluster of 
$i$ particles simultaneously located at the point $x=0$. The "integral" 
well depth for the cluster of $i$ particles we denote as
$$
{\cal U}(i)=i\, U^{(i)}\equiv i\, f(i).
$$

It is very straightforward to derive the recursion relations describing the
process under investigation. Consider the partition function (the number of
trajectories), $Z_N(x_i)$, of some arbitrary $N$--step random walk on the
halfline $x_i\in [0,\infty)$ starting at the point $x=1$ and
ending at the point $x_i$ ($x_i\in {\Z}^{+}$). The function
$Z_N(x_i)$ satisfies the equations:
\be \label{eq:3}
\left\{
\begin{array}{ll}
Z_{N+1}(x_i)=Z_N(x_i+1) + Z_N(x_i-1) & \qquad (x_i\ge 1) \\
\disp Z_{N+1}(x_i)=e^{U(x_1,\ldots,x_i,\ldots,x_n)}\, Z_N(x_i+1) &
\qquad (x_i=0)
\\
Z_{N=0}(x_i)=\delta_{x_i,0}
\end{array}
\right.
\ee

The same "equations of motion" (as Eq.(\ref{eq:3})) should be written for
all functions $Z_N(x_i)$ where $i=\in[1,n]$ and the potential
$U(x_1,\ldots,x_i,\ldots,x_n)$ is defined in Eq.(\ref{eq:1}).

We are interested in the situation when the number of particles jumping on
${\Z}^{+}$ (i.e. the number of different random walks) is very large ($n\gg 
1$) and the number of steps of each random walk tends to infinity 
($N\to\infty$).

We expect that in the thermodynamic limit $N\to \infty$ and for $n\gg 1$
the absorption transition into localized state at the point $x=0$ is
sensitive to the shape of the function $f(n)$. Namely, if $f(n)$ grows
rather slowly with $n$ (for the precise criteria of the growth see the
Section \ref{sec:3}) we could expect that the phase transitions occur in
each individual random walk independently; while if $f(n)$ is rather sharp
function of $n$, various types of {\it collective} localization are
expected. The interplay between the entropy loss and the energy gain in
the localized states makes the phase behavior of the system under
consideration extremely rich.

The most attention in our work is paid to exact determination of the
critical shape of the function $f(n)$ at which the localization transition
occurs simultaneously in all $n$ ($n\to \infty$) random walks.

To avoid further misunderstandings let us define precisely what 
we mean under the "localization" of the random walk. Let $F(N)$ be the
free energy of the $N$--step random walk on the halfline ${\Z}^{+}$
with the point well of depth $U$ at the origin $x=0$. This model can be
described via the diffusion--type equation with specific boundary conditions
(compare to Eq.(\ref{eq:3})) and there exists some critical value 
$U=U_{\rm cr}>0$ which separates two different behaviors of the free 
energy:
$$
\lim_{N\to\infty} \frac{1}{N}F(N)=\left\{\begin{array}{ll} 
0 & \mbox{for $U<U_{\rm cr}$} \\ {\rm \Lambda}>0 & \mbox{for $U>U_{\rm cr}$}
\end{array} \right.
$$
The value $U_{\rm cr}$ we call the "localization transition point". It
signals the separation of the highest eigenvalue $\Lambda$ from the 
continous spectrum of the diffusion--type equation.

\section{Exact solutions of the catalytic absorption model for
$n=\{1,\;2\}$}
\label{sec:2}

\subsection{Solution for $n=1$}
\label{sec:2.1}
\setcounter{equation}{0}

Let us start with the simplest situation when $n=1$, i.e. we have a single
simplest random walk on ${\Z}^{+}$ interacting with the potential well at
the origin.

In this case Eqs.(\ref{eq:1})--(\ref{eq:3}) degenerate to
the following ones:
\be \label{eq:4}
U(x)=\left\{
\begin{array}{ll}
0 & \mbox{if $x\neq 0$} \\
U^{(1)} & \mbox{if $x=0$}
\end{array} \right.
\ee
and
\be \label{eq:5}
\left\{
\begin{array}{ll}
Z_{N+1}(x)=Z_N(x+1) + Z_N(x-1) & \qquad (x\ge 1) \\
\disp Z_{N+1}(x)=e^{U(x)}\, Z_N(x+1) & \qquad (x=0) \\
Z_N(x)=0 & \qquad (x<0) \\
Z_{N=0}(x)=\delta_{x,0}
\end{array}
\right.
\ee
Thus, for $n=1$ we have only one particle jumping on the halfline and the
problem is nothing else as a simple absorption at the "point well" located
at the origin $x=0$.

The solution of the one--particle problem is very straightforward. Let us
make the shift $x\to x+1$. Now the equations (\ref{eq:5}) can be rewritten
as:
\be \label{eq:5.1}
\disp Z_{N+1}(x)=Z_N(x-1)+Z_N(x+1)+\delta_{x,1}
\left(e^{U^{(1)}}-1\right) Z_N(x+1) \qquad (x\ge 1)
\ee
completed by the boundary $Z_N(x=0)=0$ and initial
$Z_{N=0}(x)=\delta_{x,1}$ conditions.

Perform the substitution
\be \label{eq:5.0}
Z_N(x)=2^N\, {\cal Z}_N(x)
\ee
and use the Fourier--Laplace transform
$$
\tilde{{\cal Z}}(q,s)=\sum_{N=0}^{\infty} s^N\sum_{x=0}^{\infty} \sin
\frac{\pi x q}{l}\, {\cal Z}_N(x)
$$
Introducing the new variable $\disp k=\frac{\pi q}{l}$, we arrive
at the following integral equation for the function
$\tilde{{\cal Z}}_N(k,s)$:
\be \label{eq:5.2}
\frac{1}{s}\,\tilde{{\cal Z}}(k,s)-\frac{1}{s}\sin k = \cos k\,
\tilde{{\cal Z}}(k,s)+
\sin k\, \left(e^{U^{(1)}}-1\right)\frac{1}{\pi} \int_0^{\pi} dk
\sin 2k\,\tilde{{\cal Z}}(k,s)
\ee
After simple algebra we get
\be \label{eq:5.3}
{\cal Z}(x=2,s)\equiv \frac{1}{\pi} \int_0^{\pi} dk \sin 2k\,
\tilde{{\cal Z}}(k,s)=\frac{\disp \frac{1}{\pi}\int_0^{\pi}dk\,
\frac{\sin k\,\sin 2k}{1-s \cos k}}{\disp 1-s\left(e^{U^{(1)}}-1\right)
\frac{1}{\pi}\int_0^{\pi}dk\,\frac{\sin k\,\sin 2k}{1-s \cos k}}
\ee
The divergence of the function $\disp \frac{1}{\pi} \int_0^{\pi} dk
\sin 2k\,\tilde{{\cal Z}}(k,s)$ occurs at the critical value $U^{(1)}=
U_{\rm cr}^{(1)}$ (corresponding to the localization transition point) when 
the denominator in Eq.(\ref{eq:5.3}) is set to zero:
\be \label{eq:5.4}
e^{U_{\rm cr}^{(1)}}-1=\left.\frac{1}{\disp s\, \frac{1}{\pi}
\int_0^{\pi}dk\, \frac{\sin k\,\sin 2k} {1-s \cos k}}\right|_{s\to 1}
\to 1
\ee
Thus, in the thermodynamic limit $N\to \infty$ (i.e. when $s\to 1$) we have
$U_{\rm cr}^{(1)}=\ln 2$.

\subsection{Solution for $n=2$}
\label{sec:2.2}

\subsubsection{General ansatz}
\label{sec:2.2.1}
Consider now two particles jumping on the halfline, i.e. two trajectories
of random walks simultaneously developing in "time" $t$ ($0\le t\le N$).
These trajectories are independent on each others in all points
$x=\{1,\,2,\,\ldots\}$ except the point $x=0$. Let us call the current
coordinates of 1st and 2nd random walks $x_1(t)$ and $x_2(t)$
correspondingly. The potential (Eq.(\ref{eq:1})) reads now:
\be \label{eq:6}
U\left(x_1,x_2\right)=
\left\{
\begin{array}{ll}
0 & \mbox{if $x_1\neq 0$ and $x_2\neq 0$} \\
U^{(1)} & \mbox{if $\{x_1=0$ and $x_2\neq 0\}$ or $\{x_1\neq 0$ and
$x_2=0\}$} \\ U^{(2)} & \mbox{if $x_1=0$ and $x_2=0$} \end{array}
\right.
\ee

Thus we have the 2D random walk in the first quarter of the
$(x_1,x_2)$--plane with specific boundary conditions. In order to
simplify the corresponding equations, it is very convenient to choose the
embedding lattice as it is shown in Fig.\ref{fig:2}. The equations for the
joint partition function, $Z_N(x_1,x_2)$, in 2D--case (for $n=2$) become
more tricky than in 1D--case (for $n=1$).

Performing the shift $x_1\to x_1+1$ and $x_2\to x_2+1$ (see
Fig.\ref{fig:3}), we can derive the recursion relation for the partition
function $Z_N(x_1,x_2)$ which is the two--dimensional extension of
Eq.(\ref{eq:5.1}):
\be \label{eq:7}
\begin{array}{lll}
Z_{N+1}(x_1,x_2) & = & \disp \Delta_{x_1,x_2} Z_N(x_1,x_2)+ \\
& & \disp \delta_{x_1,1} \left(e^{U^{(1)}}-1\right)
\Delta_{x_2}Z_N(x_1+1,x_2)+ \\
& & \disp \delta_{x_2,1} \left(e^{U^{(1)}}-1\right)
\Delta_{x_1}Z_N(x_1,x_2+1)+ \\ & &
\disp \delta_{x_1,1} \delta_{x_2,1}
\left(e^{2U^{(2)}}-2e^{U^{(1)}}+1\right) Z_N(x_1+1,x_2+1)
\end{array}
\ee
where
\be \label{eq:7.0}
\Delta_{x_i}\Psi(x_i)\equiv \Psi(x_i-1)+\Psi(x_i+1) \qquad (i=\{1,2\})
\ee
and
\be \label{eq:7.01}
\Delta_{x_1,x_2}\Psi(x_1,x_2)\equiv \Psi(x_1-1,x_2-1) + \Psi(x_1-1,x_2+1) +
\Psi(x_1+1,x_2-1) + \Psi(x_1+1,x_2+1)
\ee

The equation (\ref{eq:7}) is valid for $x_1\ge 1,\; x_2\ge 1$ and should
be completed by the boundary and initial conditions:
\be \label{eq:7.1}
\left\{
\begin{array}{l}
Z_N(x_1=0,x_2\ge 0)=Z_N(x_1\ge 0,x_2=0)=Z_N(x_1=0,x_2=0)=0 \\
Z_{N=0}(x_1,x_2)=\delta_{x_1,1}\,\delta_{x_2,1}
\end{array}
\right.
\ee

It is easy to check that Eqs.(\ref{eq:7})--(\ref{eq:7.1}) describe
properly our model (taking into account the shift $x_{1,2}\to x_{1,2}+1$):

(i) For $x_1\ge 2$ and $x_2\ge 2$ we have a simple diffusion without any
interactions;

(ii) For $x_1=1$ and $x_2\ge 2$ (or for $x_1\ge 2$ and $x_2=1$) the
Boltzman weight $\beta^{(1)}$ of the random walk steps located at the point
$x_1=1$ (or $x_2=1$) reads
\be \label{eq:7.2}
\beta^{(1)}=\left(e^{U^{(1)}}-1\right)+1=e^{U^{(1)}};
\ee

(iii) For $x_1=1$ and $x_2=1$ the Boltzman weight $\beta^{(2)}$ of the 
steps of two different random walk simultaneously located at the point 
$(x_1=1,x_2=1)$ reads 
\be \label{eq:7.3}
\beta^{(2)}=\left(e^{2U^{(2)}}-2e^{U^{(1)}}+1\right)+
2\left(e^{U^{(1)}}-1\right)+1=\left(e^{U^{(2)}}\right)^2.
\ee

Let us search the solution of Eqs.(\ref{eq:7})--(\ref{eq:7.1}) using the
following ansatz
\be \label{eq:8}
Z_N(x_1,x_2)=Z_N(x_1)\,Z_N^{\rm free}(x_2)+Z_N(x_2)\,Z_N^{\rm free}(x_1)+
W_N(x_1,x_2)
\ee
where:

\noindent a) The "one--particle" partition function $Z_N(x_i)$
($i=\{1,2\}$) is the same as in Eq.(\ref{eq:5.1});

\noindent b) The term $Z_N^{\rm free}(x)$ is defined as follows
\be \label{eq:8free}
\left\{\begin{array}{ll}
Z_N^{\rm free}(x)=\Delta_x\, Z_N^{\rm free}(x) & (x\ge 1) \\
Z_N^{\rm free}(x)=0 & (x=0) \\
Z_{N=0}^{\rm free}(x)=\delta_{x,1}
\end{array} \right.
\ee
and has the sense of "free" partition function without any potential on the
halfline $x>0$;

\noindent c) The last term $W_N(x_1,x_2)$ describes the "absorption at the
{\it main corner}" and has the "nonmultiplicative nature" being determined
by the recursion relations:
\be \label{eq:9}
W_{N+1}(x_1,x_2)=\Delta_{x_1,x_2} W_N(x_1,x_2)+
\delta_{x_1,1}\,\delta_{x_2,1} \gamma_2 W_N(x_1+1,x_2+1)
\ee
where
\be \label{eq:9.1}
\left\{\begin{array}{l}
W_N(x_1=1,x_2\ge 2)=W_N(x_1\ge 2,x_2=1)=W_N(x_1\le 0,x_2\le 0)=0
\\ W_{N=0}(x_1,x_2)=\delta_{x_1,1}\,\delta_{x_2,1}
\end{array} \right.
\ee
and the "main corner" Boltzmann weight
\be \label{eq:9.2}
\gamma_2=e^{2U^{(2)}}-2e^{U^{(1)}}+1
\ee
is chosen such that Eq.(\ref{eq:8}) reproduces the right weight
$\beta^{(2)}$ at the point $(x_1=0,x_2=0)$ (see Eqs.(\ref{eq:7}) and
(\ref{eq:7.3})).

The equations (\ref{eq:9})--(\ref{eq:9.1}) correspond to the situation
shown in Fig.\ref{fig:4}a, where we marked the point at which the Boltzmann
weight $\gamma_2$ is located by the sign \unitlength=1mm \put(1.20,-1.20)
{\framebox(4.00,4.00)[cc] {$\bullet$}} \hspace{0.5cm}~.

As it has been stated in Section \ref{sec:1} we are interested in the
determination of the phase transition point of the function
$Z_N(x_1,x_2)$ (for $n=2$ and $N\gg 1$). Generalising the arguments
of Section \ref{sec:2.1} it is easy to verify that the transition to
the localized state occurs at the divergence point of the function
$Z(x_1=2,x_2=2,s)$ (as in Eq.(\ref{eq:5.3})):
\be \label{eq:10}
{\cal Z}_{N\to \infty}(x_1=2,x_2=2)=
\frac{1}{\pi^2}\int_0^{\pi}\int_0^{\pi}dk_1\,dk_2 \sin 2k_1\, \sin 2k_2\,
\tilde{{\cal Z}}(k_1,k_2,s\to 1)
\ee
where
\be \label{eq:9.0}
Z_N(x_1,x_2)=4^N\, {\cal Z}_N(x_1,x_2); \quad
W_N(x_1,x_2)=4^N\, {\cal W}_N(x_1,x_2)
\ee
and
$$
\tilde{{\cal Z}}(k_1,k_2,s)=\sum_{N=0}^{\infty}s^N \sum_{x_1=0}^{\infty}
\sum_{x_2=0}^{\infty} \sin x_1k_1\, \sin x_2k_2\, {\cal Z}_N(x_1,x_2)
$$
(compare to Eq.(\ref{eq:5.0})).

Substituting the ansatz (\ref{eq:8}) into (\ref{eq:10}) we see that the
divergence of the function $Z_{N\to\infty}(x_1=2,x_2=2)$ is determined:

$\bullet$ either by the divergence of the "one--particle" function
$Z_{N\to\infty}(x_i=2)$ ($i=\{1,2\}$)

$\bullet$ or by the divergence of the "main corner" part
$W_{N\to\infty}(x_1=2,x_2=2)$

\noindent and everything depends on the fact which of these two functions
diverges first (i.e. for smaller values of $\{U^{(1)}$ and $U^{(2)}\}$) in 
the phase space of parameters $\{U^{(1)},\, U^{(2)}\}$ (recall that 
according to Eq.(\ref{eq:1.1}) we have a restriction $0\le U^{(1)} \le 
U^{(2)}$).

The behavior of the function $Z_N(x_i)$ for $i=\{1,2\}$ has been
well studied in Section \ref{sec:2.1}; so, let us concentrate the
efforts on the solution of Eqs.(\ref{eq:9})--(\ref{eq:9.1}).

\subsubsection{Phase transition point of the "main corner" part
$W_N(x_1,x_2)$}
\label{sec:2.2.2}

The partition function $W(N)\equiv W_N(x_1=1,x_2=1)$ where $N$ is even, it is
possible to represent in the following form. Let us introduce the auxiliary
functions:
\begin{enumerate}
\item $\Omega(N)\equiv \Omega_N(x_1=1,x_2=1)$---the number of closed
paths, starting at the point ($x_1=1,x_2=1$), finishing at the same point
after $N$ steps {\it for the first time} and satisfying the boundary
conditions in the Fig.\ref{fig:4}a;
\item $V(N)\equiv V_N(x_1=2,x_2=2)$---the number of $N$--step
closed paths, starting and finishing at the point ($x_1=2,x_2=2$) and
satisfying the boundary conditions in the Fig.\ref{fig:4}b;
\end{enumerate}

The following identity can be derived immediately:
\be \label{eq:11}
\Omega(N+2)=V(N)\equiv 4^N{\cal V}(N)
\ee

The function $W(N)$ admits the representation
\be \label{eq:12}
W(N)=\sum_{k=1}^{N/2}(\beta^{(2)})^{k}\,
\sum_{\{N_1+\ldots+N_{k}=N\}}\;\prod_{i=1}^{k} \Omega(N_i)
\ee
where the upper limit in the first sum can be set to infinity because the
condition $N_1+\ldots+N_{k}=N$ ensures the right cut of the sum;
$\beta^{(2)}=\gamma_2+1$ is the Boltzmann weight of the potential well at
the origin.

Introducing the Kronecker $\delta$--function:
$$
\delta(x)=\frac{1}{2\pi i}\oint\frac{dz}{z^{1+x}}=
\left\{\begin{array}{ll} 0 & x\neq 0 \\ 1 & x=0
\end{array}\right.
$$
where $x=N-N_1-\ldots-N_k$, we may rewrite Eq.(\ref{eq:12}) as follows:
\be \label{eq:13}
\begin{array}{lll}
W(N) & = & \disp \sum_{k=0}^{\infty} \left(\beta^{(2)}\right)^k\,
\frac{1}{2\pi i} \oint dz\,z^{-N-1} \sum_{N_1=2}^{\infty}\ldots
\sum_{N_k=2}^{\infty} z^{N_1+\ldots+N_k} \prod_{i=1}^k \Omega(N_i) \\
& = & \disp \frac{1}{2\pi i} \oint dz\,z^{-N-1} \frac{1}{\disp 1-
\beta^{(2)}\,\sum_{N=2}^{\infty}z^N \Omega(N)}
\end{array}
\ee

The appearance of the pole in the last expression signals the separation of
the localized mode from the continuous part of the spectrum corresponding
to the function $W(N)$. Using the identity (\ref{eq:11}) we get the
equation on the transition point (for $N\to \infty$):
\be \label{eq:14}
\beta_{\rm cr}^{(2)} = \lim_{z\to 4}\;
\frac{1}{\disp \sum_{N=0}^{\infty}\left(\frac{z}{4}\right)^N {\cal V}(N)}
\equiv \lim_{s\to 1}\; \frac{1}{\disp \sum_{N=0}^{\infty} s^N {\cal V}(N)}
\ee
where $\disp s=\frac{z}{4}$.

The function $V_N(x_1,x_2)$ satisfies the recursion relation in
absence of any potentials:
\be \label{eq:15}
V_{N+1}(x_1,x_2) = \Delta_{x_1,x_2}\, V_N(x_1,x_2)
\ee
with the boundary conditions shown in Fig.\ref{fig:4}b:
\be \label{eq:15.1}
\left\{
\begin{array}{l}
V_N(x_1=1,x_2\ge 2)=V_N(x_1\ge 2,x_2=1)= V_N(x_1=1,x_2=1)=0
\\ V_{N=0}(x_1,x_2)=\delta_{x_1,2}\,\delta_{x_2,2}
\end{array}
\right.
\ee

Using the 2D--Fourier and Laplace transforms and performing the back shift
in Eq.(\ref{eq:15}) $\{x_1\to x_1-1,\, x_2\to x_2-1\}$ we arrive at the
standard equation for the function ${\cal V}_N(k_1,k_2,s)$:
\be \label{eq:16}
{\cal V}_N(k_1,k_2,s)= \frac{\sin k_1 \sin k_2}{1-s \cos k_1 \cos k_2}
\ee

Thus, we get the final expression:
\be \label{eq:17}
\sum_{N=0}^{\infty} s^N {\cal V}(N)=\frac{1}{\pi^2} \int\limits_0^{\pi}
\int\limits_0^{\pi}dk_1\,dk_2\; \frac{\sin^2 k_1
\sin^2 k_2}{1-s \cos k_1 \cos k_2}
\ee
Evaluating the last integral at the point $s=1$ we obtain the following
numerical value for the "main corner" transition point:
\be \label{eq:18}
\beta_{\rm cr}^{(2)}\equiv \gamma_2+1=\frac{\pi}{4-\pi} \approx 3.65979
\ee

\subsubsection{Phase diagram for 2--particle system}
\label{sec:2.2.3}

Collecting the results of the Sections \ref{sec:2.2.1} and \ref{sec:2.2.2}
we conclude that the localization transition from the delocalized to the
absorbed state is determined:
\begin{itemize}
\item either by the equation $\beta_{\rm cr}^{(1)} \equiv
e^{U_{\rm cr}^{(1)}}=2$ (the "one--particle" part);
\item or by the equation $\disp \beta_{\rm cr}^{(2)}=\gamma_2+1$, i.e.
$e^{2U^{(2)}}-2e^{U^{(1)}}+2=\frac{\pi}{4-\pi}$ (the two--particle "main
corner" part).
\end{itemize}

The complete phase diagram is drawn in Fig.\ref{fig:5}. For $U^{(1)}>\ln 2$
the influence of the potential $U^{(2)}$ disappears, while for
$0<U^{(1)}<\ln 2\approx 0.693$ the "induced" localization can occur
for
$$
U^{(2)}> \frac{1}{2}\ln\left(2e^{U^{(1)}}+
\frac{\pi}{4-\pi}-2\right)
$$
just due to the simultaneous interactions between two random
walks at the origin.

It is worthwhile to mention that the "collective" localization can appear
even for $U^{(1)}<0$. Suppose, for example, such situation:
$$
U^{(1)}=-\infty; \quad U^{(2)}>0
$$
i.e. each particular random walk cannot visit the origin, but there is an
energy gain for two random walks to stay at the origin simultaneously. In
this case for
$$
U^{(2)}>U^{(2)}_{\rm cr}=\frac{1}{2}\ln\left(\frac{\pi}{4-\pi}-2\right)
\approx 0.253
$$
the trajectories become localized.

\section{Phase transitions in catalytic absorption model for $n>2$}
\label{sec:3}
\setcounter{equation}{0}

\subsection{Solution for $n=3$}
\label{sec:3.1}

The equation for the 3--particle partition function $Z_N(x_1,x_2,x_3)$
reads:
\be \label{eq:19}
\begin{array}{lll}
Z_N(x_1,x_2,x_3) & = & \Delta_{x_1,x_2,x_3}Z_N(x_1,x_2,x_3)+ \\
& & \hspace{-1cm} \delta_{x_1,1}\left(e^{U^{(1)}}-1\right)
\Delta_{x_2,x_3} Z_N(x_1+1,x_2,x_3)+ \\
& & \hspace{-1cm} \delta_{x_2,1}\left(e^{U^{(1)}}-1\right)
\Delta_{x_1,x_3} Z_N(x_1,x_2+1,x_3)+ \\
& & \hspace{-1cm} \delta_{x_3,1}\left(e^{U^{(1)}}-1\right)
\Delta_{x_1,x_2} Z_N(x_1,x_2,x_3+1)+ \\
& & \hspace{-1cm} \delta_{x_1,1} \delta_{x_2,1}\left(e^{2U^{(2)}}-
2e^{U^{(1)}}+1\right) \Delta_{x_3}Z_N(x_1+1,x_2+1,x_3)+ \\
& & \hspace{-1cm} \delta_{x_1,1} \delta_{x_3,1}\left(e^{2U^{(2)}}-
2e^{U^{(1)}}+1\right) \Delta_{x_2}Z_N(x_1+1,x_2,x_3+1)+ \\
& & \hspace{-1cm} \delta_{x_2,1} \delta_{x_3,1}\left(e^{2U^{(2)}}-
2e^{U^{(1)}}+1\right) \Delta_{x_1}Z_N(x_1,x_2+1,x_3+1)+ \\
& & \hspace{-1cm} \delta_{x_1,1} \delta_{x_2,1} \delta_{x_3,1}
\left(e^{3U^{(3)}}-3e^{2U^{(2)}}+3e^{U^{(1)}}-1\right)
Z_N(x_1+1,x_2+1,x_3+1)
\end{array}
\ee

Generalising the ansatz (\ref{eq:8}) to the 3D--case we may search the
solution of Eq.(\ref{eq:19}) in the form:
\be \label{eq:20}
\begin{array}{lll}
Z_N(x_1,x_2,x_3) & = & Z_N(x_1)Z_N^{\rm free}(x_2,x_3)+
Z_N(x_2)Z_N^{\rm free}(x_1,x_3)+ Z_N(x_3)Z_N^{\rm free}(x_1,x_2)+ \\
& & W_N(x_1,x_2)Z_N^{\rm free}(x_3)+W_N(x_1,x_3)Z_N^{\rm free}(x_2)+
W_N(x_2,x_3)Z_N^{\rm free}(x_1)+ \\
& & W_N(x_1,x_2,x_3)
\end{array}
\ee
where:

\noindent a) The partition function $W_N(x_i,x_j)$ ($i\neq j;\; \{i,j\}\in
[1,3]$) is the same as in Eq.(\ref{eq:9});

\noindent b) The functions $Z_N^{\rm free}(x_i)$ and $Z_N^{\rm
free}(x_i,x_j)$ ($i\neq j;\; \{i,j\}\in [1,3]$) describe the contributions
of one-- and two--dimensional free random walks with zero's boundary
conditions (compare to Eq.(\ref{eq:8}));

\noindent c) The function $W_N(x_1,x_2,x_3)$ is determined as follows:
\be \label{eq:20w}
\left\{\begin{array}{l}
W_{N+1}(x_1,x_2,x_3)=\Delta_{x_1,x_2,x_3} W_N(x_1,x_2,x_3)+
\delta_{x_1,1}\,\delta_{x_2,1} \delta_{x_3,1} \gamma_3
W_N(x_1+1,x_2+1,x_3+1) \\
W_N(x_1=1,x_2\ge 2,x_3\ge 2)=W_N(x_1\ge 2,x_2=1,x_3\ge 2)=
W_N(x_1\ge 2,x_2\ge 2,x_3=1)=0 \\
W_N(x_1=1,x_2=1,x_3\ge 2)=W_N(x_1=1,x_2\ge 2,x_3=1)=
W_N(x_1\ge 2,x_2=1,x_3=1)=0 \\
W_{N=0}(x_1,x_2)=\delta_{x_1,1}\,\delta_{x_2,1}
\end{array} \right.
\ee
with $\gamma_3=e^{3U^{(3)}}-3e^{2U^{(2)}}+3e^{U^{(1)}}-1$.

In order to check the validity of ansatz (\ref{eq:20}) let us compute the
values of the Boltzmann weights at the boundaries:

1. {\it At the planes:}
\be \label{eq:21a}
\left. \begin{array}{ll}
x_1=0 & (x_2>0,x_3>0) \\
x_2=0 & (x_1>0,x_3>0) \\
x_3=0 & (x_1>0,x_2>0)
\end{array} \right\}:
\quad \beta^{(1)}=\left(e^{U^{(1)}}-1\right)+1=e^{U^{(2)}}
\ee

2. {\it At the lines:}
\be \label{eq:21b}
\left. \begin{array}{ll}
(x_1=0, x_2=0) & x_3>0 \\
(x_1=0, x_3=0) & x_2>0 \\
(x_2=0, x_3=0) & x_1>0
\end{array} \right\}:
\quad
\begin{array}{lll}
\beta^{(2)}&=&e^{2U^{(2)}}-2e^{U^{(1)}}+1+ \\
&& 2\left(e^{U^{(1)}}-1\right)+1= e^{2U^{(2)}} \end{array}
\ee

3. {\it At the "main corner" point:}
\be \label{eq:21c}
(x_1=0,x_2=0,x_3=0): \quad
\begin{array}{lll}
\beta^{(3)}&=&e^{3U^{(3)}}-3e^{2U^{(2)}}+3e^{U^{(1)}}-1-
3\left(e^{U^{(1)}}-1\right)- \\ &&
3\left(e^{2U^{(2)}}-2e^{U^{(1)}}+1\right)+1=e^{3U^{(3)}}
\end{array}
\ee

The arguments of the Section \ref{sec:2.2.2} can be easily extended to the
3D--case. Skipping the intermediate computations we bring the final result
for the value of the Boltzmann weight at the "main corner" transition
point:
\be \label{eq:24}
\beta_{\rm cr}^{(3)}=\left(\frac{1}{\pi^3} \int\limits_0^{\pi}
\int\limits_0^{\pi}\int\limits_0^{\pi}dk_1\,dk_2\,dk_3\; \frac{\sin^2 k_1
\sin^2 k_2 \sin^2 k_3}{1- \cos k_1 \cos k_2 \cos k_3}\right)^{-1}
\approx 7.856
\ee

The localization phase transition is now determined:
\begin{itemize}
\item either by the equation $\beta_{\rm cr}^{(1)}=2$, i.e.
\be \label{eq:25a}
U_{\rm cr}^{(1)}=\ln 2
\ee
(the one--particle part);
\item or by the equation $\disp \beta_{\rm cr}^{(2)}=\gamma_2+1=
\frac{\pi}{4-\pi}$, i.e.
\be \label{eq:25b}
U_{\rm cr}^{(2)}=\frac{1}{2} \ln\left(2e^{U^{(1)}}+\frac{\pi}{4-\pi}-2
\right)
\ee
(the two--particle "main corner" part);
\item or by the equation $\disp \beta_{\rm cr}^{(3)}=\gamma_3+1=
7.856$, i.e.
\be \label{eq:25c}
U^{(3)}_{\rm cr}=\frac{1}{3} \ln \left(7.856+
3e^{2U^{(2)}}-3e^{U^{(1)}}\right)
\ee
(the three--particle "main corner" part).
\end{itemize}

\subsection{Solution for arbitrary $n$}
\label{sec:3.2}

The general ansatz for arbitrary values of $n$ reads
\be \label{eq:26a}
\begin{array}{lll}
Z_N(x_1,x_2,\ldots,x_n) & = & Z_N(x_1)Z_N^{\rm free}(x_2,\ldots,x_n)+
\mbox{all permutations}+\\
& & \hspace{-1cm} W_N(x_1,x_2)Z_N^{\rm free}(x_3,\ldots,x_n)+
\mbox{all permutations}+ \\
& & \hspace{-1cm} W_N(x_1,x_2,x_3)Z_N^{\rm free}(x_4,\ldots,x_n)+
\mbox{all permutations} + \\ & & \cdots + \\
& & \hspace{-1cm} W_N(x_1,\ldots,x_{n-1})Z_N^{\rm free}(x_n)
+\mbox{all permutations} + \\
& & \hspace{-1cm} W_N(x_1,\ldots,x_n)
\end{array}
\ee
Eq.(\ref{eq:26a}) produces the following recursive relation for the value
$\gamma_n$:
\be \label{eq:26b}
\gamma_n=e^{nU^{(n)}}-C_n^1\gamma_{n-1}-C_n^2\gamma_{n-2}-C_n^3
\gamma_{n-3}-\ldots
\ee
completed by the initial condition $\gamma_0=1$.

The localization transition in the $n$--particles model occurs at the
values $\beta^{(j)}_{\rm cr}$ ($j=2,3,4,\ldots,n$) defined by the equation
\be \label{eq:28}
\beta^{(j)}_{\rm cr}=\gamma_j+1=\left(\frac{1}{\pi^j}
\int\limits_0^{\pi}\cdots\int\limits_0^{\pi} dk_1\,\cdots\,dk_j\;
\frac{\sin^2 k_1\cdots \sin^2 k_j}{1- \cos k_1
\cdots \cos k_j}\right)^{-1}
\ee
where $\gamma_j$ are the recursive solutions of Eq.(\ref{eq:26b}).

\subsection{Phase transition in the $n$--particle system for $n \gg 1$}
\label{sec:3.3}

Now we are in position to give the answer to the question raised in 
Section \ref{sec:1}: "What should be the critical shape of the function 
$f(n)$ (for $n\gg 1$) to have the simultaneous collective $n$--particle
localization in the system?" For simplicity later on we suppose $n$ to be
even. The answer is very straightforward: {\it The condition on the critical
shape $f_{\rm cr}(n)$ providing the joint $n$--particle localization is
defined by setting $U^{(n)}_{\rm cr} \equiv f_{\rm cr}(n)$ in the equation
{\rm (\ref{eq:26b})} with $\gamma_j$ extracted from {\rm (\ref{eq:28})}}.

The approximate evaluation of the integral (\ref{eq:28}) for $n\gg 1$ gives
us (with the exponential accuracy---see Appendix for details):
\be \label{eq:29}
\gamma_n=\exp\left(\frac{3}{2}n\ln n -\frac{1}{2}n \ln
\frac{\pi e^3}{54}+O(\ln n) \right)
\ee
We drop in Eq.(\ref{eq:28}) all terms growing with $n\to \infty$ slower
than $\exp({\rm const}\;n)$.

Substituting (\ref{eq:28}) into (\ref{eq:26a}) it is easy to conclude that
for $n\gg 1$ the term $\gamma_n$ has the maximal value in the sum
(\ref{eq:26a}):
\be \label{eq:30}
\gamma_n = e^{\frac{3}{2}n \ln n}
\ee
where we kept the leading asymptotics (for $n\to \infty$) only.

Estimating the sum (\ref{eq:26a}) with the same accuracy, we arrive at the
following final conclusion:
\begin{itemize}
\item If the function $f(n)$ grows faster than $f_{\rm cr}(n)=\frac{3}{2}
\ln n$ then {\bf all $n\gg 1$ particles are localized
simultaneously} in the point well located at the origin $x=0$ of the
halfline ${\Z}^{+}$.
\end{itemize}

Recall that $U^{(n)}$ is the depth of just the $n$--particle well. Thus the
"integral" critical depth, ${\cal U}_{\rm cr}(n)=n U_{\rm cr}^{(n)}$ per
the cluster of $n$ particles (i.e. steps of different random walks at the
point $x=0$) has the following asymptotic behavior
$$
{\cal U}_{\rm cr}(n)= \frac{3}{2} n \ln n
$$

If the integral potential at the origin ${\cal U}(n)$ grows (with $n\to
\infty$) slower than ${\cal U}_{\rm cr}(n)=\frac{3}{2}n \ln n$ then only a
finite fraction of random walks remains at the origin in the localized
state.

\section{Instead of discussion: Connection with random heteropolymer
absorption problem}
\label{sec:4}
\setcounter{equation}{0}

The above mentioned problem has a direct application to the old problem of the
random copolymer absorption at the point well. Namely consider the random
walk on ${\Z}^{+}$ with the point well located at the origin and suppose
that the well depth depends on the current "time" (i.e. the number of the 
random walk step) at the point $x=0$. The partition function $\Theta_N$ of 
such process one can write as a sum over all random walk paths
\be \label{eq:31}
\Theta_N(x_N)={\cal N}\sum_{\{x_1,\ldots,x_{N-1}\}}\prod_{j=1}^N\left\{
g\left(|x_j-x_{j-1}|\right)e^{U_j(x_j)}
\right\}
\ee
where ${\cal N}$ is the normalization constant, $g\left(|x_j-
x_{j-1}|\right)$ is the local transition probability and the potential
$U_j(x_j)$ is the random variable of $j$. Suppose that $U_j(x_j)$ allows
the following representation:
\be \label{eq:32}
U_j(x_j)=
\left\{\begin{array}{ll}
\frac{1+\sigma_j}{2}\,\varepsilon_1+\frac{1-\sigma_j}{2}\,\varepsilon_2
& \mbox{for $x_j=0$} \\ 0 & \mbox{for $x_j>0$} \end{array} \right.
\ee
where $\varepsilon_{1,2}$ are some positive constants and $\sigma_j$ is the
random variable defined as follows:
$$
\sigma_j=\left\{\begin{array}{ll} +1 & \mbox{with the probability $p$} \\
-1 & \mbox{with the probability $1-p$} \end{array} \right.
$$

Write (\ref{eq:31}) in form of recursion relations:
\be \label{eq:33}
\left\{\begin{array}{ll}
\Theta_{N+1}(x)=\Theta_N(x+1)+\Theta_N(x-1) & (x\ge 1) \\
\Theta_{N+1}(x)=e^{U_N}\Theta_N(x+1) & (x=1) \\
\Theta_N(x)=0 & (x<0)\\
\Theta_{N=0}(x)=\delta_{x,0} &
\end{array} \right.
\ee
Performing the shift $x\to x+1$ rewrite (\ref{eq:33}) as follows
\be \label{eq:34}
\left\{\begin{array}{ll}
\Theta_{N+1}(x)=\Delta_x\Theta_N(x)+\left(e^{U_N}-1\right)\delta_{x,1}
\Theta_N(x+1) & (x\ge 1) \\
\Theta_N(x)=0 & (x=0) \\
\Theta_{N=0}(x)=\delta_{x,1} &
\end{array} \right.
\ee

The partition function $\Theta_N(x)$ is the random variable which depends
on the quenched random pattern of realisations of the potential $U_j$
($j\in[1,N]$) at the origin. To find the reliable thermodynamic quantity we
have to average the free energy $\ln \Theta_N$ over the distribution of all
random sequences $\{\sigma_1,\sigma_2,\ldots,\sigma_N\}$. The corresponding
computations we realise in the frameworks of the replica approach.

Averaging the $n$s power of the partition function
$\left<\Theta_N^n(x)\right>=\Phi_N(x_1,\ldots,x_n)$, we get:
\be \label{eq:35}
\begin{array}{lll}
\Phi_{N+1}(x_1,..,x_n) & = & \disp \Delta_{x_1,..,x_n}
\Phi_N(x_1,..,x_n)+ \\ & & \disp \hspace{-3cm} \left<e^{U_N}-1\right>
\sum_{i=1}^n \delta_{x_i,1} \Delta_{x_1,..,x_i \hspace{-6pt}
\backslash,..,x_n} \Phi_N(x_1,..,x_{i-1},x_i+1,x_{i+1},..,x_n)+ \\
& & \disp \hspace{-3cm} \left<\left(e^{U_N}-1\right)^2\right> \sum_{i>j}^n
\delta_{x_i,1}\delta_{x_j,1} \Delta_{x_1,..,x_i \hspace{-6pt}
\backslash,..,x_j \hspace{-6pt} \backslash,..,x_n}\Phi_N(x_1,..,x_{i-1},
x_i+1,x_{i+1},..,x_{j-1},x_j+1,x_{j+1},..,x_n)+ \\ & & \cdots + \\
& & \disp \hspace{-3cm} \left<\left(e^{U_N}-1\right)^n\right>
\delta_{x_1,1}\ldots\delta_{x_n,1} \Phi_N(x_1+1,..,x_n+1)
\end{array}
\ee

The solution of Eq.(\ref{eq:35}) reads (compare to (\ref{eq:26a})):
\be \label{eq:36}
\begin{array}{lll}
\Phi_N(x_1,x_2,\ldots,x_n) & = & \overline{\Phi}_N(x_1)
Z_N^{\rm free}(x_2,\ldots,x_n)+\mbox{all permutations}+\\
& & \hspace{-1cm} \overline{\Phi}_N(x_1,x_2)
Z_N^{\rm free}(x_3,\ldots,x_n)+\mbox{all permutations}+ \\
& & \hspace{-1cm} \overline{\Phi}_N(x_1,x_2,x_3)
Z_N^{\rm free}(x_4,\ldots,x_n)+\mbox{all permutations} + \\
& & \cdots + \\ & & \hspace{-1cm} \overline{\Phi}_N(x_1,\ldots,x_{n-1})
Z_N^{\rm free}(x_n)+\mbox{all permutations} + \\
& & \hspace{-1cm} \overline{\Phi}_N(x_1,\ldots,x_n)
\end{array}
\ee
where the functions $\overline{\Phi}_N(x_1,\ldots,x_k)$ satisfy the
master equations ($k\in[1,n]$)
\be \label{eq:37}
\left\{\begin{array}{ll}
\overline{\Phi}_N(x_1,..,x_j)=\Delta_{x_1,..x_j}
\overline{\Phi}_N(x_1,..,x_j)+ \delta_{x_1,1}..\delta_{x_j,1}
\overline{\gamma}_j \overline{\Phi}_N(x_1+1,..,x_j+1) &
\{x_1,..x_j\}\ge 1 \\
\overline{\Phi}_N(\mbox{at least one $x_i\ge 2$ and all others
$x_j=1$})=0 & \{i\neq j\}\in[1,j] \\
\overline{\Phi}_{N=0}(x_1,..x_n)=\delta_{x_1,1}..\delta_{x_n,1} &
\end{array} \right.
\ee
and by $\overline{\gamma}_j$ we denote the averaged values of the Boltzmann
weights:
\be \label{eq:38}
\overline{\gamma}_j=\left<\left(e^{U_N}-1\right)^j\right>=
p\left(e^{\varepsilon_1}-1\right)^j+(1-p)\left(e^{\varepsilon_2}-1\right)^j
\ee

Using the results of the previous Sections one can conclude that the
averaged moments of the quenched heteropolymer partition function,
$\left<\Theta_N\right>,\;\left<\Theta_N^2\right>,\dots,
\left<\Theta_N^n\right>$ exhibit the singular behavior at the set of
points being the solutions of following equations:
\be \label{eq:39}
\beta^{(j)}_{\rm cr}=\overline{\gamma}_j+1=p\left(e^{\varepsilon_1}-1
\right)^j+(1-p)\left(e^{\varepsilon_2}-1\right)^j+1
\ee
where $\beta^{(1)}_{\rm cr}=2$ (see Eq.(\ref{eq:5.4})) and
$\beta^{(j)}_{\rm cr}$ for $j\in[2,n]$ is given by Eq.(\ref{eq:28}). Recall
that we restrict ourselves with the case: $\varepsilon_1>0$ and
$\varepsilon_2>0$.

The "true critical point" of the localization transition in all averaged
moments of the partition function $\Theta_N$ (in the thermodynamic limit
$N\to\infty$) can be obtained using the following simple procedure. We fix
some arbitrary value $\varepsilon_2$ and find the {\bf minimal} value
$\varepsilon_1^{\rm cr}(\varepsilon_2,j)$ among all solutions of
Eq.(\ref{eq:39}) for $j\in[1,n]$.

It is easy to check that for all $1\le j\le n$ and any arbitrary choice of
$\varepsilon_2$, the minimal value $\varepsilon_1^{\rm cr}$ corresponds 
just to $j=1$.  It means that {\it all moments of the random copolymer 
partition function, $\Theta_N^j$, averaged over the quenched disorder in 
monomer types diverge at the same point as the "one--particle" part (i.e.
"annealed" copolymer partition function), $\left<\Theta_N\right>$.}

Of course, our consideration has an obvious crucial shortcoming connected
with the fact that the replica approach presented above does not allow us
to take properly the limit $n\to 0$. Thus, the computations performed in
Section \ref{sec:4} cannot be regarded as a proof of the conjecture
that the phase transition points of copolymers with quenched and annealed
chemical sequences coincide. However, the consistency of our investigation
with other speculations on that subject \cite{gro2,forg} at least gives
hope that our conclusion is correct.

\subsection*{Acknowledgement}
I am very grateful to G. Uimin, S. Korshunov and A. Grosberg for useful
suggestions concerning the model itself and for critical comments regarding
its possible application to the heteropolymer problem. Also I would thank
H. Spohn for his conjectures about the critical shape of the potential 
$f(n)$ and for drawing my attention to the Refs.\cite{krug,evans}.  I 
highly appreciate valuable discussions of the problem with G. Oshanin, A. 
Comtet, S.A. Gredeskul and V. Tchijov.

\newpage

\begin{appendix}
\section{Appendix}
\setcounter{equation}{0}

Let us estimate the value of the integral $I_n=(\beta^{(n)})^{-1}$ (see
Eq.(\ref{eq:24})) for $n\gg 1$
\be \label{eq:a1}
I_n=\frac{1}{\pi^n}\int_0^{\pi}\ldots\int_0^{\pi} dk_1\ldots dk_n \,
\frac{\sin^2 k_1\ldots \sin^2 k_n}{1-\cos k_1\ldots \cos k_n}
\ee
Changing the variables $k_i=q\pi$ and expanding the nominator and
denominator of the fraction of (\ref{eq:a1}) up to the first non vanishing
term, we get:
\be \label{eq:a2}
I_n\approx\frac{1}{2^{n-1}}\int_{-1}^1\ldots\int_{-1}^1 dk_1\ldots dk_n\,
\frac{k_1^2\ldots k_n^2}{k_1^2+\ldots+k_n^2}
\ee
Passing to the $n$--dimensional spherical coordinate system we arrive after
simple algebra at the following expression:
\be \label{eq:a3}
I_n=\frac{(\sqrt{\pi})^{n-1}}{3n-2}\prod_{l=1}^{n-1}\left\{
\frac{1}{9 l^2-1}\frac{\Gamma\left(\frac{3l}{2}\right)}
{\Gamma\left(\frac{3l-1}{2}\right)}\right\}
\ee

In the limit $n\gg 1$ we find the asymptotic expression of the function
$I_n$ with the exponential accuracy:
\be \label{eq:a4}
I_n=\exp\left(-\frac{3}{2}n\ln n +\frac{1}{2}n \ln
\frac{\pi e^3}{54}+O(\ln n) \right)
\ee
This asymptotic expression has been used in the derivation of the equation
(\ref{eq:30}).

\end{appendix}

\newpage

\section*{Figure Captions}
\bigskip

\begin{fig}
The schematic representation of the multiparticle absorption model.
\label{fig:1}
\end{fig}
\bigskip

\begin{fig}
The choise of the lattice for the 2D--random walk in the first quater
$(x_1>0,\,x_2>0)$ and the possible local moves of the walker.
\label{fig:2}
\end{fig}
\bigskip

\begin{fig}
The lattice $(x_1>1,\,x_2>1)$ with zero's boundary conditions obtained from
the lattice shown in Fig.\ref{fig:2} by the shift $x_{1,2}\to x_{1,2}+1$.
\label{fig:3}
\end{fig}
\bigskip

\begin{fig}
(a) The auxiliary 2D--lattice used in the computation of the
nonmultiplicative contribution $W_N(x_1,x_2)$---see Eq.(\ref{eq:9}); (b)
The same lattice as in (a) but with another boundary condition at the "main
corner".
\label{fig:4}
\end{fig}
\bigskip

\begin{fig}
The phase diagram for the two--particle catalytic absorption model.
\label{fig:5}
\end{fig}

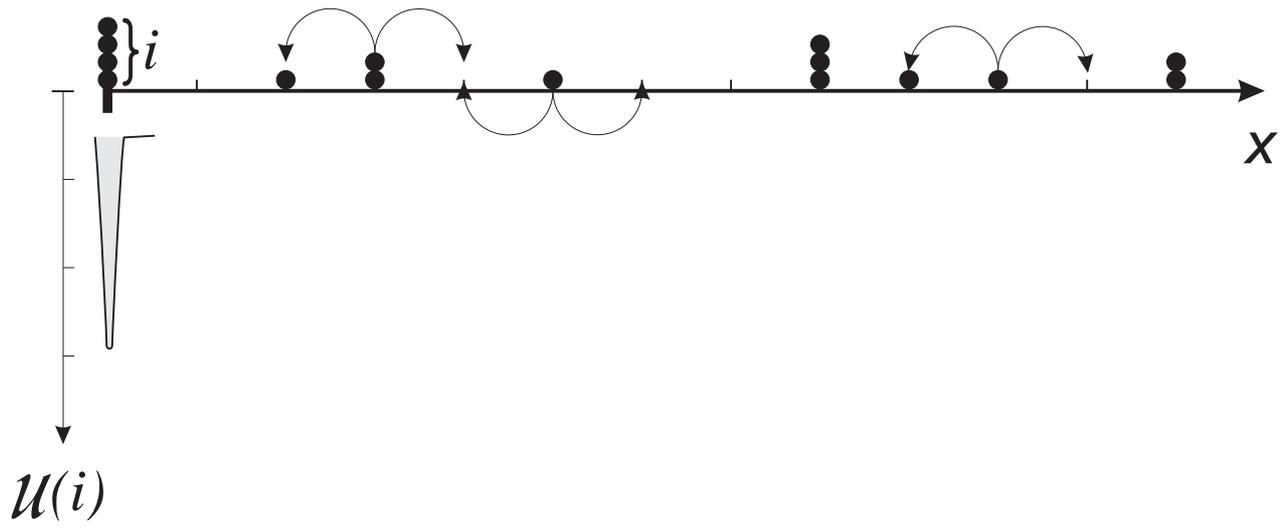
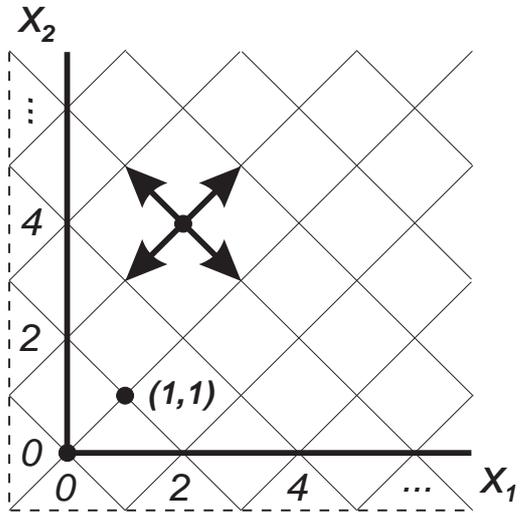
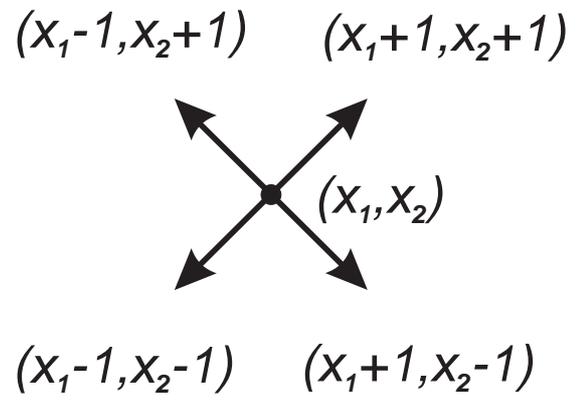
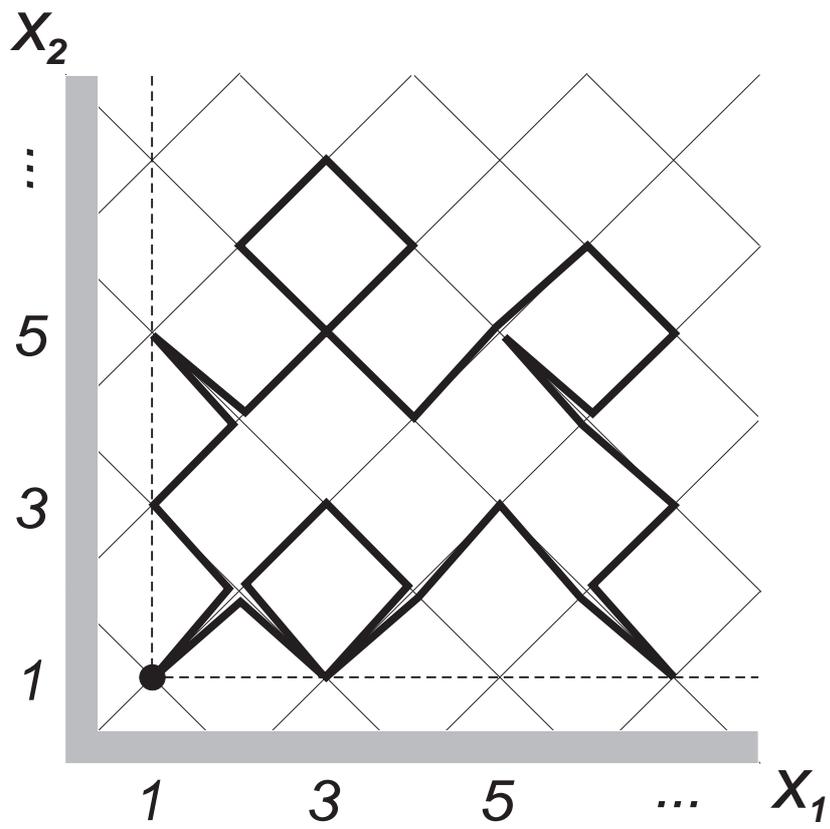
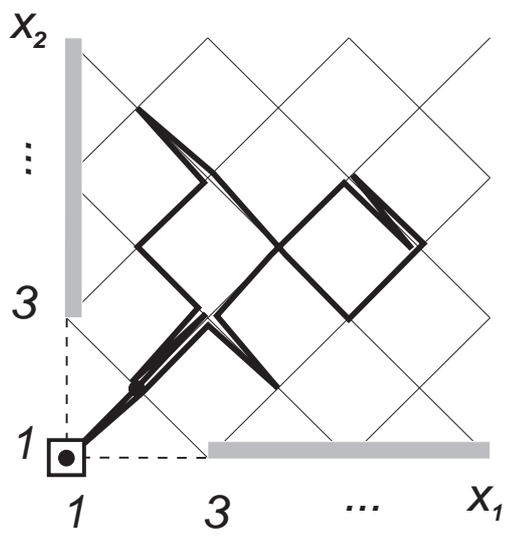
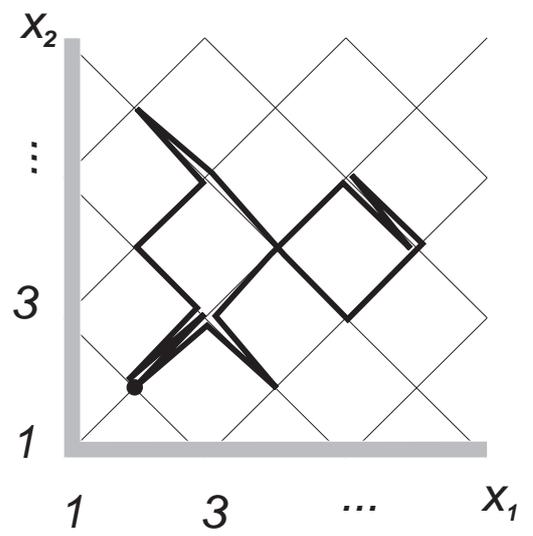
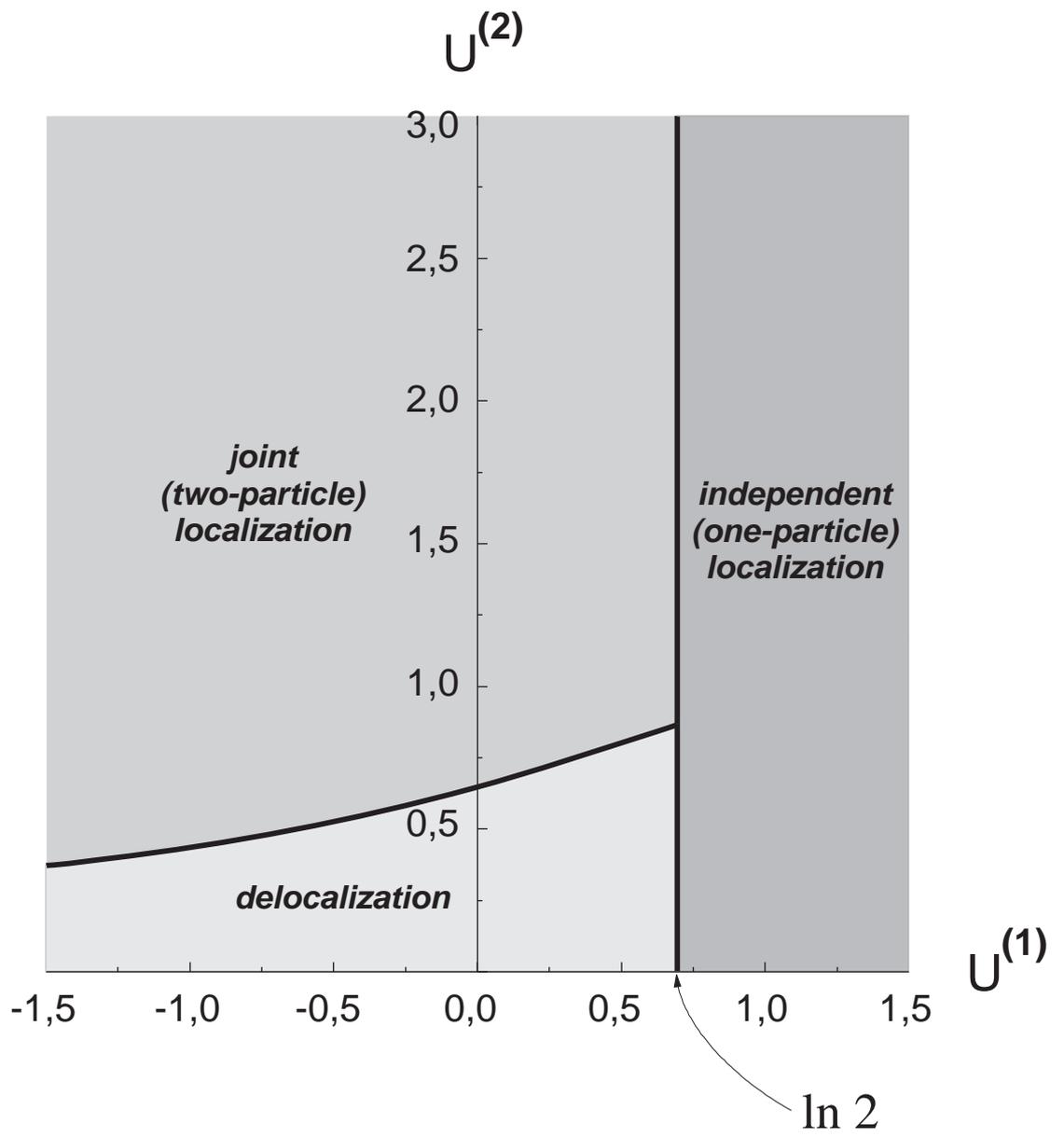
\end{document}